\documentclass[twocolumn,showpacs,prl,nofootinbib,superscriptaddress,preprintnumbers,amsmath,amssymb,floatfix]{revtex4}
\usepackage{graphicx}
\usepackage{dcolumn}
\usepackage{bm}

\begin{document}
\title{Fermi LAT Search for Photon Lines from 30 to 200 GeV \\ and Dark Matter Implications}

\author{A.~A.~Abdo}
\affiliation{Space Science Division, Naval Research Laboratory, Washington, DC 20375, USA}
\affiliation{National Research Council Research Associate, National Academy of Sciences, Washington, DC 20001, USA}
\author{M.~Ackermann}
\affiliation{W. W. Hansen Experimental Physics Laboratory, Kavli Institute for Particle Astrophysics and Cosmology, Department of Physics and SLAC National Accelerator Laboratory, Stanford University, Stanford, CA 94305, USA}
\author{M.~Ajello}
\affiliation{W. W. Hansen Experimental Physics Laboratory, Kavli Institute for Particle Astrophysics and Cosmology, Department of Physics and SLAC National Accelerator Laboratory, Stanford University, Stanford, CA 94305, USA}
\author{W.~B.~Atwood}
\affiliation{Santa Cruz Institute for Particle Physics, Department of Physics and Department of Astronomy and Astrophysics, University of California at Santa Cruz, Santa Cruz, CA 95064, USA}
\author{L.~Baldini}
\affiliation{Istituto Nazionale di Fisica Nucleare, Sezione di Pisa, I-56127 Pisa, Italy}
\author{J.~Ballet}
\affiliation{Laboratoire AIM, CEA-IRFU/CNRS/Universit\'e Paris Diderot, Service d'Astrophysique, CEA Saclay, 91191 Gif sur Yvette, France}
\author{G.~Barbiellini}
\affiliation{Istituto Nazionale di Fisica Nucleare, Sezione di Trieste, I-34127 Trieste, Italy}
\affiliation{Dipartimento di Fisica, Universit\`a di Trieste, I-34127 Trieste, Italy}
\author{D.~Bastieri}
\affiliation{Istituto Nazionale di Fisica Nucleare, Sezione di Padova, I-35131 Padova, Italy}
\affiliation{Dipartimento di Fisica ``G. Galilei", Universit\`a di Padova, I-35131 Padova, Italy}
\author{K.~Bechtol}
\affiliation{W. W. Hansen Experimental Physics Laboratory, Kavli Institute for Particle Astrophysics and Cosmology, Department of Physics and SLAC National Accelerator Laboratory, Stanford University, Stanford, CA 94305, USA}
\author{R.~Bellazzini}
\affiliation{Istituto Nazionale di Fisica Nucleare, Sezione di Pisa, I-56127 Pisa, Italy}
% ...

\date{\today}
\begin{abstract}
Dark matter (DM) particle annihilation or decay can produce monochromatic $\gamma$-rays readily distinguishable from astrophysical sources. $\gamma$-ray line limits from 30 GeV to 200 GeV obtained from 11 months of Fermi Large Area Space Telescope data from 20-300 GeV are presented using a selection based on requirements for a $\gamma$-ray line analysis, and integrated over most of the sky.  We obtain $\gamma$-ray line flux upper limits in the range $0.6-4.5\times 10^{-9}\mathrm{cm}^{-2}\mathrm{s}^{-1}$, and give corresponding DM annihilation cross-section and decay lifetime limits. Theoretical implications are briefly discussed.


\end{abstract}

\preprint{}
 \maketitle


\emph{Introduction:} Annihilation or decay of dark matter (DM) particles can give rise to Standard Model particles, including $\gamma$-rays.
The predicted $\gamma$-ray flux is usually weak compared to astrophysical sources, and striking experimental signatures are valuable in recognizing a signal. A distinctive signal would be the detection of a narrow $\gamma$-ray line
\cite{Srednicki:1985sf,Rudaz:1986db,Bergstrom:1988fp,Rudaz:1990rt,Jungman1996, Bergstrom1997, Bergstrom1998}
originating from dark matter particles annihilating or decaying into $\gamma X$, where $X$ can be another photon, a $Z$-boson, a Higgs-boson, or a non-Standard Model particle.
A $\gamma$-ray line search using data from EGRET in the energy range
$1-10$ GeV has been presented in \cite{Pullen2007}.
We present here for the first time results from 11 months of Fermi Large Area Telescope (LAT) data in the energy range \mbox{20-300 GeV} and obtain limits for \mbox{30-200 GeV}.

DM particles of mass $m_\chi$ annihilating into $\gamma X$ produce monochromatic $\gamma$-rays of energy
$E_\gamma = m_\chi \Big(1 - \frac{m_X^2}{4m_\chi^2}\Big)$, where $m_X$ is the mass of $X$.
For decays $\chi\to \gamma X$, the equation gives the photon energy after the substitution $m_{\chi}\to m_{\chi}/2$.
Since dark matter is strongly constrained to be electrically neutral, it has no direct coupling to photons.
The process $\chi(\chi) \to \gamma X$ thus occurs only at higher orders, and
with a branching fraction that is typically only $10^{-4}-10^{-1}$ compared to the total annihilation or decay rate.
However, a strong line feature, or a strong line-like feature such as a sharp cut-off at a particular energy, can be
expected in models with supersymmetry
\cite{Bergstrom:1997fh,Bern:1997ng,Ullio:1997ke,Bergstrom:2005ss,Bringmann2008}, an extended Higgs sector
\cite{Gustafsson2007}, from final state radiation
\cite{Beacom:2004pe,Bergstrom:2004cy,Birkedal2005},  from gravitino decay
\cite{Ibarra:2007wg}, or in scenarios with non-thermal WIMP production \cite{Kane:2009if}.

\emph{LAT Data Selection and Analysis:}
The LAT is a pair-conversion telescope that combines silicon-strip/tungsten trackers and hodoscopic
CsI(Tl) calorimeters into a 4x4 array of 16 identical modules. The tracker is covered by a segmented anti-coincidence detector (ACD). Including the tracker, the LAT presents 10 radiation lengths for normal incidence. The depth, segmentation, and wide field of view of the LAT enable its high-energy reach. Separation of the very large charged cosmic-ray background from $\gamma$-rays is achieved via the combination of the data acquisition trigger, on-board event software filter system, and extensive ground processing of the data. Details of the LAT, and the data analysis are given in \cite{Atwood2009}. An account of how the LAT is calibrated on orbit is presented in \cite{Eduardo2009}. The LAT nominally operates in a scanning mode that covers the whole sky every two orbits ($\sim 3$ hours). The analysis described here uses data taken in this scanning mode from Aug 7, 2008, to July 21, 2009, corresponding to an average exposure of $3.3\times10^{10}$ cm$^{2}$s.

Events are selected for this analysis only if they pass additional cuts to the Pass6V3 diffuse class cuts, i.e., the cleanest photon sample in the currently public Fermi data release \cite{Atwood2009, publicfermi}. These additional cuts are: a) a small average charge deposited in the tracker planes (veto against heavy ions); b) a transverse shower size in the calorimeter within a size range expected for electromagnetic showers (veto against hadronic showers and minimum ionizing particles). Cuts a) and b) dramatically reduce the charged particle background at the loss of some effective area, yielding a $\gamma$-ray efficiency $> 90\%$ relative to that of Pass6V3 diffuse class. These cuts remove charged particle backgrounds that would worsen our upper limits, and that can even yield structures that might be interpreted as $\gamma$-ray lines. These cuts are a subset of those used in the LAT Measurements of the Isotropic Diffuse Gamma-Ray Emission\cite{EGB}.

In addition, we use only one of the three energy measurement methods from the LAT standard analysis. The method used is the LAT profile method \cite{Atwood2009}, in order to not introduce energy dependent artifacts that arise from switching between methods over the energy range discussed here. In the profile method, the energy is obtained from a fit to the longitudinal shower profile while considering the transverse profile. The exclusive use of the profile energy method led to no additional reduction in efficiency. The instrument response functions (IRFs) are recalculated for this data selection and for the use of the profile method.

The resulting energy resolution averaged over the LAT acceptance is 11\% FWHM  for 20-100 GeV, increasing to 13\% FWHM for 150-200 GeV. The photon angular resolution is less than $0.1^{\circ}$ over the energy range of interest ($68\%$ containment). The absolute calibration and energy resolution of the LAT was determined by comparing with $e^-$ beam test data, taken at CERN in a secondary $e^-$ beam with energies up to 300 GeV using a special calibration unit made up of flight spare towers and ACD tiles (not the LAT itself) \cite{FermiCRE2009}. The energy resolution measured in the beam tests is in agreement with the predictions from the Monte Carlo simulator based on GEANT4 that was used to define the IRFs (GLEAM \cite{Atwood2009}).  Also, the systematic error on the absolute energy of the LAT was determined to be $-10 +5$\% for 20-300 GeV.

The systematic uncertainties for the exposure over this energy range are $\pm 20\%$ based on the extrapolation of studies comparing the efficiencies of analysis cuts for data and simulation of observations of Vela \cite{FermiVela2008}. The Vela studies cut off at 10 GeV, and measurements of the systematic errors above 10 GeV will be made when sufficient statistics are available from high-energy sources. Thus, the exposure systematic errors above 10 GeV that we quote for this study have not been fully validated. We believe that any reasonably projected uncertainty would not have a significant impact on the interpretation of the limits presented here.

In our search for lines we use a region of interest (ROI) that covers most of the sky: an all-sky ROI with the Galactic plane removed (i.e., $|b|>10^\circ$ as the Galactic plane is very bright in photons from gas interactions) \emph{plus} a $20^\circ \times 20^\circ$ square centered on the Galactic center (GC) and aligned on the $(\ell,b)$ grid of the Galactic coordinate system (Galactic coordinates in degrees are used in this paper). Though increasing the photon flux limits averaged over the reported energy range by less than $10\%$, including the GC gives significantly better theoretical line limits. For the highly point source rich region within $1^\circ$ of the GC, no point source removal was done as this would have removed the GC. For the remaining part of the ROI, point sources obtained from the year-1 catalog under development by the LAT team \cite{LATCAT2009} were masked from the analysis using a circle of radius $0.2^\circ$ centered on the measured point source position (conservative, considering the angular resolution above 20 GeV). This last cut removes about 0.4\% of the solid angle and about 5\% of the total photons.

\begin{figure}[t!]
\begin{center}
\includegraphics[width=0.90\columnwidth,angle=0]{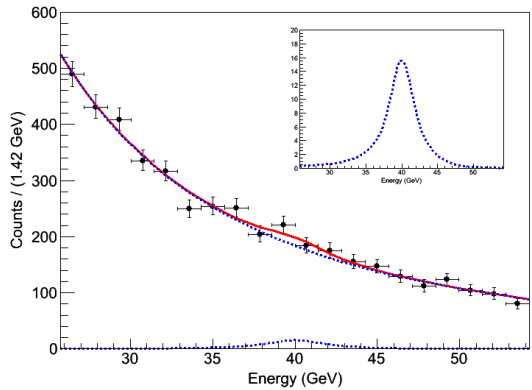}
\end{center}
\caption{A binned representation of a typical fit (done unbinned), here centered at 40 GeV, used to extract the flux upper limits presented in the tables. The fitting process is described in the text. In the main part of the figure, the lower (upper) dotted line is the signal (background) from the fit and the red(or black) line is the total fit. This fit also shows the largest line ''signal" that was obtained in the reported energy region. The inset shows a blow-up of the signal, which is the line energy response function, $S(E)$, used in this fit, and is typical of line shapes for 20-300 GeV. \label{fig:fit}}
\end{figure}

In searching for deviations from a locally-determined background, we use a sliding energy window with the window size adjusted to reflect the energy resolution. This results in limits for $30<E_\gamma<200$ GeV using data in the range $20<E_\gamma<300$ GeV.  The response  of the LAT to a line feature in energy were determined from full detector simulations (GLEAM \cite{Atwood2009}).
Fig.~\ref{fig:fit} shows a binned representation of the fit and a close-up of the LAT line response function at 40 GeV. This fit also shows the largest line ''signal" that was obtained in the reported energy region.  We use an unbinned likelihood method, with the likelihood function $L\left(\bar{E}|f,\Gamma\right) = \prod\limits_{i=0}^{n_{tot}} f\cdot S\left(E_{i}\right) + \left(1-f\right)\cdot B\left(E_{i},\Gamma\right)$.
Here, $E_{i}$ denotes the measured energy of the $i^{th}$ photon. The parameters  $f$ and $\Gamma$ are free and represent the signal fraction and the index of the power-law function,$B\left(E_{i},\Gamma\right)$, used to model the background. We require $f \geq 0$ in the fit. The function $S\left(E_{i}\right)$ models the signal shape, i.e., the LAT response for a line feature in energy averaged over the acceptance of the LAT. The confidence intervals are determined using the profile likelihood method (MINOS within MINUIT)\cite{MINUIT}, which provides two sided confidence intervals.  The properties of this statistical method (coverage and power) have been thoroughly tested. At 100 GeV, for example, the coverage is close to nominal for a range of true signal fractions from 0 to 50\%, and the power reaches 100\% for signal fractions of about 10\%. The method overcovers slightly due to the physical constraint on the signal fraction, $f$.

\begin{table*}[t]
\begin{center}
\begin{ruledtabular}
\begin{tabular}{c||c|lll||c|lll}
$E_\gamma$ & $95\%$CLUL & \multicolumn{3}{c||}{$\langle\sigma v\rangle_{\gamma\gamma}$ [$\gamma Z$]  ($10^{-27}$ cm$^{3}$s$^{-1}$)} & \multicolumn{3}{c}{$\tau_{\gamma\gamma}$ [$\gamma Z$]  ($10^{28}$ s)} \\
(GeV) & ($10^{-9}$ cm$^{-2}$s$^{-1}$) & NFW & Einasto & Isothermal & NFW & Einasto & Isothermal \\
\hline
30 &  3.5 & 0.3 [2.6] & 0.2 [1.9] & 0.5 [4.5] & 17.6 [4.2] & 17.8 [4.2] & 17.5 [4.2] \\
40 &  4.5 & 0.7 [4.2] & 0.5 [3.0] & 1.2 [7.2] & 10.1 [2.9] & 10.3 [2.9] & 10.0 [2.9] \\
50 &  2.4 & 0.6 [2.7] & 0.4 [1.9] & 1.0 [4.6] & 15.5 [5.0] & 15.7 [5.1] & 15.4 [5.0] \\
60 &  3.1 & 1.1 [4.2] & 0.8 [3.0] & 1.8 [7.3] & 9.8 [3.5] & 10.0 [3.5] & 9.7 [3.5] \\
70 &  1.2 & 0.6 [2.0] & 0.4 [1.4] & 1.0 [3.4] & 21.6 [8.2] & 21.9 [8.3] & 21.5 [8.1] \\
80 &  0.9 & 0.5 [1.7] & 0.4 [1.2] & 0.9 [2.9] & 26.0 [10.4] & 26.4 [10.5] & 25.8 [10.3] \\
90 &  2.6 & 2.0 [6.0] & 1.5 [4.3] & 3.5 [10.3] & 7.7 [3.2] & 7.8 [3.2] & 7.6 [3.1] \\
100 &  1.4 & 1.4 [3.8] & 1.0 [2.8] & 2.4 [6.6] & 12.6 [5.4] & 12.8 [5.4] & 12.5 [5.3] \\
110 &  0.9 & 1.0 [2.7] & 0.7 [1.9] & 1.7 [4.6] & 18.9 [8.2] & 19.2 [8.3] & 18.8 [8.2] \\
120 &  1.1 & 1.6 [4.0] & 1.1 [2.9] & 2.7 [6.9] & 13.3 [5.9] & 13.5 [6.0] & 13.2 [5.9] \\
130 &  1.8 & 3.0 [7.3] & 2.1 [5.3] & 5.1 [12.6] & 7.6 [3.4] & 7.8 [3.5] & 7.6 [3.4] \\
140 &  1.9 & 3.5 [8.4] & 2.5 [6.0] & 6.0 [14.3] & 7.0 [3.2] & 7.1 [3.3] & 7.0 [3.2] \\
150 &  1.6 & 3.5 [8.2] & 2.5 [5.9] & 6.0 [14.1] & 7.5 [3.5] & 7.6 [3.5] & 7.4 [3.4] \\
160 &  1.1 & 2.7 [6.3] & 2.0 [4.5] & 4.7 [10.9] & 10.2 [4.8] & 10.4 [4.8] & 10.1 [4.7] \\
170 &  0.6 & 1.7 [4.0] & 1.3 [2.9] & 3.0 [6.8] & 17.0 [8.0] & 17.2 [8.1] & 16.9 [7.9] \\
180 &  0.9 & 2.7 [6.1] & 1.9 [4.4] & 4.6 [10.4] & 11.6 [5.5] & 11.8 [5.6] & 11.6 [5.4] \\
190 &  0.9 & 3.2 [7.1] & 2.3 [5.1] & 5.5 [12.2] & 10.4 [4.9] & 10.5 [5.0] & 10.3 [4.9] \\
200 &  0.9 & 3.3 [7.3] & 2.4 [5.2] & 5.7 [12.5] & 10.6 [5.1] & 10.8 [5.1] & 10.5 [5.0] \
\end{tabular}
\caption{Flux, annihilation cross-section upper limits, and decay lifetime lower limits: $\gamma$-ray energies measured and  corresponding $95\%$ c.l.~upper limits (CLUL) on fluxes, for $|b|>10^\circ$ \emph{plus} a $20^\circ \times 20^\circ$ square around the Galactic center. For each energy and flux limit, $\langle\sigma v\rangle_{\gamma\gamma}$  and $\langle\sigma v\rangle_{\gamma Z}$ upper limits, and $\tau_{\gamma\gamma}$ and $\tau_{\gamma Z}$ lower limits are given for three Galactic dark matter distributions (see text). The systematic error in the absolute energy of the LAT discussed in the text propagates to a  $-20\% +10\%$ systematic error on $\langle\sigma v\rangle_{\gamma\gamma}$, while for  the decay lower limits the systematic error in the absolute energy of the LAT discussed in the text propagates to a $+10\% -5\%$ systematic error on $\tau_{\gamma\gamma}$.}
\label{tab:CrossSections}
\end{ruledtabular}
\end{center}
\end{table*}


\emph{Results \& Discussion:}
Table \ref{tab:CrossSections} shows flux limits as a function of photon energy that can be translated into bounds on the annihilation cross-section
or decay lifetime assuming a specific halo dark matter density profile,
$\rho(\vec{r})$.
The monochromatic gamma-ray flux from dark matter annihilating into $\gamma X$ with a cross-section $\langle\sigma v\rangle$ is $\Phi = \frac{N_\gamma}{8\pi} \frac{\langle\sigma v\rangle}{m_\chi^2}\,\mathcal{L}$, where $N_\gamma = 2$ for $X=\gamma$ and $N_\gamma=1$ otherwise. Here,
\begin{equation}\label{eq:L}
\mathcal{L} = \int db \int d\ell \int ds\, \cos b \,\rho^2(\vec{r}),
\end{equation}
where the integral is over the ROI, $r = (s^2 + R_{\odot}^2 - 2 s R_\odot \cos\ell \cos b)^{1/2}$, and
 $R_\odot \simeq 8.5$ kpc is the distance from the sun to the GC \cite{GCDistance}.
For decays,  the flux is given by substituting in the equation for $\Phi$,
$\langle\sigma v\rangle/2m_{\chi}^2 \to 1/\tau m_\chi$,
where $\tau$ is the DM lifetime, and $\rho^2\to \rho$ in Eq.~(\ref{eq:L}) for $\mathcal{L}$.

We consider three theoretically-motivated halo profiles:  the NFW profile,
$\rho_{\rm NFW}(r) = \rho_s/[(r/r_s)(1+r/r_s)^2]$
with $r_s=20$ kpc  \cite{Navarro:1996gj}, the Einasto profile,
$\rho_{\rm Einasto}(r) = \rho_s \exp\{ -(2/\alpha)[(r/r_s)^\alpha - 1]\}$
with $r_s=20$ kpc and $\alpha=0.17$~\cite{Einasto:1965,Navarro:2008kc},
and the very shallow isothermal profile
$\rho_{\rm isothermal}(r) = \rho_s/(1+(r/r_s)^2)$
with $r_s=5$ kpc \cite{Bahcall:1980fb}.  We determine $\rho_s$ using $\rho(R_\odot)=0.4$ GeV cm$^{-3}$
\cite{Catena:2009mf}.
Taking the mass of the Milky-Way halo to be $\sim 1.2 \times 10^{12} M_\odot$ (see e.g.~\cite{Wilkinson:1999hf,Xue:2008se}),
we determine maximum values for $r$ of $\sim 150$ kpc for the Einasto and NFW profiles, and \mbox{$\sim 100$ kpc} for the isothermal profile.

Table \ref{tab:CrossSections} shows the
cross-section and lifetime limits for the above profiles. We verified that there is only a minor dependence of the flux upper limits when changing the lower bound of $|b|$ in the range $8^\circ < |b| < 15^\circ$, leaving the GC region fixed.
The cross-section limits are sensitive to the halo profile. For the ROI used here, DM annihilation has a value of $\mathcal{L}$ (Eq.~(\ref{eq:L})) for the Einasto profile that is ~40\% larger than for the NFW profile, while for DM decays the $\mathcal{L}$ values are almost the same. This sensitivity is greater for cuspier profiles than those discussed here. For example the Moore density profile \cite{Moore:1997sg} gives a factor of $\sim 6$ stronger limits than the Einasto profile, using lower bounds on $\ell$ and $b$ for the Moore profile integration that correspond to a distance of $10^{-3}$ pc from the Galactic center.

The limits on $\langle\sigma v\rangle_{\gamma\gamma}$ ($\langle\sigma v\rangle_{\gamma Z}$)
shown in Table \ref{tab:CrossSections} are about one or more orders of magnitude weaker than
the cross-sections expected for a typical thermal WIMP. However, there are several models in
the literature that predict larger cross-sections and are constrained by these results.
A WIMP produced non-thermally may have a much larger annihilation cross-section than
a thermally produced WIMP and still produce the required DM relic density.
An example is the ``Wino LSP'' model \cite{Kane:2009if} that explains the recent positron measurement by PAMELA \cite{Adriani2008}, and predicts $\langle \sigma v \rangle_{\gamma Z} \simeq 1.4 \times 10^{-26}$ cm$^3$ s$^{-1}$ at $E_\gamma \sim 170$ GeV.
Our results disfavor this model by about a factor of $\sim 2-5$, depending on the dark matter halo profile (see Table \ref{tab:CrossSections}). Other models that are partially constrained include \cite{Mambrini:2009ad},
while models that are only constrained assuming a much cuspier profile include \cite{Bertone:2009cb,Bringmann:2007nk}.

Dark matter decays into $\gamma X$  are ruled out for lifetimes below $\sim 10^{29}$s, a limit that is largely independent of the dark
matter halo profile. This constrains, for example, a subset of the lifetime range of interest for gravitinos decaying into mono-energetic photons \cite{Ibarra:2007wg}.

\emph{Acknowledgements:} We thank Louis Lyons for very useful discussions. The \textit{Fermi} LAT Collaboration
acknowledges generous ongoing support
from a number of agencies and institutes that have supported both the
development and the operation of the LAT as well as scientific data analysis. These include the National Aeronautics and Space Administration and the Department of Energy in the United States, the Commissariat \`a l'Energie Atomique and the Centre National de la Recherche Scientifique / Institut National de Physique Nucl\'eaire et de Physique des Particules in France, the Agenzia Spaziale Italiana and the Istituto Nazionale di Fisica Nucleare in Italy, the Ministry of Education, Culture, Sports, Science and Technology (MEXT), High Energy Accelerator Research Organization (KEK) and Japan Aerospace Exploration Agency (JAXA) in Japan, and the K.~A.~Wallenberg Foundation, the Swedish Research Council and the Swedish National Space Board in Sweden. Support for science analysis during the operations phase is gratefully acknowledged from the Istituto Nazionale di Astrofisica in Italy and the Centre National d'\'Etudes Spatiales in France.

\bibliography{WIMPLinesNew090909}

\end{document}